# FURTHER DEVELOPMENTS IN GENERATING TYPE-SAFE MESSAGING

R. Neswold, C. King

FNAL[†], Batavia, IL 60510, U.S.A.

*Abstract*

At ICALEPCS '09, we introduced a source code generator that allows processes to communicate safely using data types native to each host language[1]. In this paper, we discuss further development that has occurred since the conference in Kobe, Japan, including the addition of three more client languages, an optimization in network packet size and the addition of a new protocol data type.

## INTRODUCTION

Writing software to support external messaging can be tedious and error-prone. The software author must make sure the required message fields are present, are aligned properly, and follow an agreed upon byte-ordering. If he cuts any corners while implementing the protocol, problems will appear – sometimes weeks or month later. Decoding packets takes extra care since the software must prove the sender's data is well-formed; in addition to byte-ordering and alignment issues, all fields must be validated and range-checked.

The protocol compiler is a command-line utility that generates the source code that safely serializes messages. The compiler takes, as input, a source file that describes the messages found in the protocol and creates the source files that implement it.

The benefits of using the compiler rather than hand-coding the protocol are:

1. Endian issues are handled automatically so the programmer only works with the native data types of their machine.
2. Packing issues are handled automatically so message fields are aligned properly.
3. The messages are implemented in a form that is native to the language (structs in C++, classes in Java, etc.)
4. Messages are completely validated (field values are range-checked, enumerations are validated, all required fields are present.)

Over the past two years, we've used the protocol compiler on many new projects. It has allowed us to focus all our attention on the application design because we know the communication layer just works. Despite it being a very useful tool, we have found a few minor shortcomings. The solutions to which are described in this paper.

```
enum Status { Away, Online };

request SendMessage {
    string text;
}

request UpdateStatus {
    Status newStatus;
}

reply NewMessage {
    string from;
    string text;
}

reply NewStatus {
    string who;
    Status status;
}
```

Figure 1: Protocol source used for examples

## ADDING ENUMERATION TYPES

While designing protocols, we noticed a particular idiom repeatedly being used. We would use an `int16` type to represent a field that could only have a limited set of values (the very use-case that justified adding enumerations to many programming languages.) Emulating these enumerations required us to write our own validation code to make sure the values sent and received were members of the limited set. But type-safety and field validation is exactly what the protocol compiler is supposed to do! So enumerated types were deemed necessary and added as an official data type.

In the generated code of each target language, the underlying integer for each enumeration value is based on a hash value, instead of being sequential, to prevent software from assuming any ordering of the values (and, hence, breaking later if the enumeration is changed.) Programmers are encouraged to ignore the underlying integer values and simply use the symbols provided.

## NEW SUPPORTED LANGUAGES

Although the officially supported languages at Fermilab are C++ and Java, there are many groups that use alternate languages to write useful code that isn't meant for the Control Room operators. We've supported these efforts mostly by routing their data requests through Java servlets using the XMLRPC protocol. This approach works, but limits the clients' data rates and adds another "hop" to the return path. If we were to support these additional languages using the protocol compiler, they could request the data directly from the control system.

---

Table 1: Protocol Type to Language Type Mapping

| Protocol Type | C++ | Java | Python | Erlang | Objective-C |
|---|---|---|---|---|---|
| `bool` | `bool` | `bool` | `True` or `False` | atoms `'true'` or `'false'` | `NSNumber*` |
| `int16` | `int16_t` | `short` | `int` | integer with restricted range | `NSNumber*` |
| `int32` | `int32_t` | `int` | `int` | integer with restricted range | `NSNumber*` |
| `int64` | `int64_t` | `long` | `long` | integer with restricted range | `NSNumber*` |
| `double` | `double` | `double` | `float` | `float` | `NSNumber*` |
| `string` | `std::string` | `string` | `str` | `string` (which, in Erlang, is a list of integers) | `NSString*` |
| `binary` | `std::vector<uint8_t>` | `byte[]` | `bytearray` | `binary` | `NSData*` |
| `struct T {}` | `struct T {}` | `class T {}` | `class T` | tuple defined with a record specification | Objective-C class |
| `enum T {}` | `enum T {}` | `enum T {}` | `int` with restricted values | atoms | `enum T {}` |
| `optional T` | `std::auto_ptr<T>` | nil object references | presence or absence of class attributes | atom `'nil'` or actual value | `nil` or actual value |
| `T[]` | `std::vector<T>` | `T[]` | list of T | list of T | `NSArray*` of objects |

Table 1 shows the five currently supported languages and how the protocol data types map to their respective native data types.

The examples given in the following sections assume the protocol input file, `example.proto` (shown in Figure 1), was used.

*Python Programming Language*

There is a community of Python programmers at Fermilab which lean heavily on the XMLRPC protocol, so it made sense to add Python as target language.

Python's data types are less numerous than there are in C++, so the Python integer type serves several purposes. The generated code makes sure the values are in range before encoding or decoding messages. In addition, Python is a dynamically typed language. So, even though the generated code creates "official" protocol classes to be used for the messages, any class with the correct set of attributes can be used (attributes not defined in the protocol will not be part of the serialization.)

We chose to use the Python iterator interface as the mechanism for marshalling and unmarshalling messages. The unmarshal routines accept and read from a character iterator to decode the message while the marshal routines return an iterator which generates the character stream.

Supporting the example protocol in Figure 1 using Python first requires creating the module that serializes the messages. The following command creates the `example.py` source file:

```
pc -l python example.proto
```

After importing this new module in a Python script, a `SendMessage` message can be marshalled thusly:

```
msg = SendMessage_request()
msg.text = "Hello"
iter = marshal_request(msg)
```

The resulting iterator can be passed to anything that uses an iterator to obtain content: file objects; sockets; and even the bytearray constructor, if simply marshalling to a memory buffer is desired.

Unmarshalling requires a character iterator holding the incoming stream of data:

```
msg = unmarshal_reply(iter)
if isinstance(msg, NewMessage_reply) then:
    print msg.from, ": ", msg.text
```

The resulting message will be stored in `msg`. The specific type of message can be determined using `isinstance()`. If a message cannot be encoded or decoded, a `ProtocolError` exception is raised.

*Objective-C*

One line of mobile devices uses Java as its preferred development language, which we already support. But there is another line of popular, mobile products we also intend to use, so we added Objective-C as a target. It should be noted that this isn't a generic Objective-C generator; it uses objects defined in the OSX and iOS foundation frameworks and is, therefore, Macintosh- and iOS-specific.

The compiler generates two source files: a `.h` file containing the object interfaces and a `.m` file containing the implementations. All message fields are represented by pointers to Cocoa objects (`NSNumber*`, `NSString*`, `NSArray*`, etc.), which follows coding conventions used in Macintosh and iOS development. Required fields in a message will always point to a valid Cocoa object and optional fields will either have a valid pointer or be set to `nil`. This generator encodes to and decodes from an `NSData*` object.

Converting the Python example to objective-C results in the following sample for marshalling a request:

```
example_SendMessage_request* msg =
    [[example_SendMessage_request alloc]
init];

msg.text = @"Hello";

NSData* bin = [msg marshal];
```

Since objective-C doesn't have namespaces or a way to nest classes, the names of the protocol messages need to carry enough information to make them unique and end up being quite long. However, objective-C code tends to be verbose, so long names don't stand out as much as they would in other languages.

To unmarshal data, the binary representation is loaded into an `NSData*` object which is then passed via the protocol's unmarshal method:

```
NSData* raw = [[NSData alloc] init];

// ...load the NSData buffer here...

example* msg =
    [[example unmarshal:raw] retain];

if ([msg isKindOfclass:
        [example_NewMessage_reply class]])
{
    example_NewMessage_reply* tmp =
        (example_NewMessage_reply*) msg;

    NSLog(@"%@: %@", [tmp.from],
[tmp.text]);
}
```

If there were any problems decoding the message, an `NSException*` is thrown.

### *Erlang*

Recently, our department has taken an interest in the soft real-time, functional language Erlang[2] and we're looking for ways to use it. Before this can happen, though, Erlang needs to be made a first-class citizen in the control system by enabling it to communicate via ACNET and enabling it send and receive protocol messages. As of 2010, both those requirements were met.

The code generated for Erlang serializes protocol messages to and from Erlang binaries[1]. A binary is read in from a file or a network socket and sent to the unmarshalling function to be converted to an Erlang data type.

---

[1]The actual implementation isn't as symmetric as it sounds because messages are encoded as an Erlang "iolist" which is a binary or a list of binaries.

Table 2: Sample Integer encodings

| Value | Encoding |
|-------|----------|
| 0     | 0x00     |
| 1     | 0x01     |
| -1    | 0xff     |
| 127   | 0x7f     |
| -128  | 0x80     |
| 128   | 0x00, 0x80 |
| -129  | 0xff, 0x7f |

The Erlang language doesn't have the concept of a pointer so, to represent a missing optional field, its value is set to the atom `nil`.

Reaching back to our example, the Erlang version that encodes a message looks like:

```
Msg =
    #example_sendmessage_request{text="Hello"
},
Bin = example:marshal_request(Msg)
```

The resulting binary can be written to a socket or file. Like objective-C, Erlang doesn't have namespaces so identifiers tend to be verbose to ensure uniqueness.

Retrieving a value from a binary is as easy. In fact, it's so simple, we'll make this code segment do a little more than the previous examples. We'll use Erlang's pattern matching to show how we can decode both request types specified in Figure 1:

```
Bin = … %% obtained from socket, file, etc.
case example:unmarshal_request(Bin) of
 #example_sendmessage_request{text=T} →
   process_message(T);
 #example_updatestatus_request{newstatus=S}
→
   process_status(S)
end
```

## OPTIMIZATIONS

Using the compiler has revealed a few opportunities for improvement. This section discusses a few techniques we used to improve both the encoded size and the performance of the encoders.

### *Packet Size Improvements*

The encoded format uses integers to represent integers values, length fields and hashed field names. These integers were typically encoded using four bytes. We found that, most of the time, the length fields of strings and binaries were much smaller than the full range we allowed, resulting in many 0 bytes getting emitted. Integers values, too, didn't always push their range, so many 0 bytes (and FF bytes for negative integers) were emitted.

To reduce the superfluous data, we redefined the way integers are encoded. The current approach treats integers as the minimum number of bytes needed to encode a signed value. Some examples of the new encoding are shown in Table 2.

Unfortunately, the new encoding is incompatible with the older method, so the version of the header was bumped to version 2 (which guarantee messages with different integer encodings will fail when unmarshalling.)

*Performance Improvements*

When we originally wrote the protocol compiler, we focused on correct behavior before tackling performance issues. Now that we're satisfied with its reliability, we took a step back to see if we could improve the generated code.

It became clear that routines that emit integer values were also being called for integers that were constant (the hash values for message fields, for instance.) We replaced those calls at run-time with code that simply emits the corresponding constant byte sequence.

## CONCLUSION

The protocol compiler is continuing to prove itself as an easy and robust way to get applications written in different languages hosted on different computer architectures to communicate. We have two active Erlang projects that are using the protocol compiler to access ACNET data at high data rates. We also used the protocol compiler output to deliver ACNET data to an iPhone/iPad application.

Since it takes an average of two weeks to support a new language, we're willing to expand the protocol compiler to support new languages that our community uses.